\documentclass[%
 reprint,
superscriptaddress,
 amsmath,amssymb,
 aps,
prapplied,
]{revtex4-2}
\usepackage{graphicx}
\usepackage{dcolumn}
\usepackage{bm}
\usepackage{hyperref}
\usepackage{xcolor}

\usepackage{comment}
\usepackage[separate-uncertainty=true]{siunitx}
\usepackage[siunitx]{circuitikz}

\begin{document}

\preprint{APS/123-QED}

\title{A network of parametrically driven silicon nitride mechanical membranes}

\author{Luis Mestre}
\affiliation{Laboratory for Solid State Physics, ETH Zürich, 8093 Zürich, Switzerland}
\affiliation{Quantum Center, ETH Zürich, 8093 Zürich, Switzerland}

\author{Suyash Singh}
\affiliation{Laboratory for Solid State Physics, ETH Zürich, 8093 Zürich, Switzerland}
\affiliation{Quantum Center, ETH Zürich, 8093 Zürich, Switzerland}

\author{Gabriel Margiani}
\affiliation{Laboratory for Solid State Physics, ETH Zürich, 8093 Zürich, Switzerland}
\affiliation{Quantum Center, ETH Zürich, 8093 Zürich, Switzerland}

\author{Letizia Catalini}
\affiliation{Laboratory for Solid State Physics, ETH Zürich, 8093 Zürich, Switzerland}
\affiliation{Quantum Center, ETH Zürich, 8093 Zürich, Switzerland}
\affiliation{Center for Nanophotonics, AMOLF, 1098XG Amsterdam, The Netherlands}

\author{Alexander Eichler}
\affiliation{Laboratory for Solid State Physics, ETH Zürich, 8093 Zürich, Switzerland}
\affiliation{Quantum Center, ETH Zürich, 8093 Zürich, Switzerland}

\author{Vincent Dumont}\thanks{Corresponding author: \href{mailto:vdumont@phys.ethz.ch}{vdumont@phys.ethz.ch}} 
\affiliation{Laboratory for Solid State Physics, ETH Zürich, 8093 Zürich, Switzerland}
\affiliation{Quantum Center, ETH Zürich, 8093 Zürich, Switzerland}

\date{\today}
\begin{abstract}
Networks of nonlinear resonators offer a promising platform for analog computing and the emulation of complex systems. However, realizing such networks remains challenging, as it requires resonators with high quality factors, individual frequency tunability, and strong inter-resonator coupling. In this work, we present a system that meets all these criteria. Our system is based on metallized silicon nitride membranes that are coupled via their common substrate and controlled capacitively via electrodes. We demonstrate individual frequency tuning and strong parametric driving of each membrane. Notably, we tune membrane frequencies through avoided crossings and demonstrate tunability of the coupled membrane's parametric response. This platform provides a scalable and controllable setting for exploring collective phenomena, dynamical phase transitions, nonlinear topology, and analog computing.
\end{abstract}

\maketitle

\textit{Introduction.}--- Networks of coupled oscillators can emulate complex dynamical systems, offering a physical platform to solve computationally hard problems and to explore emergent phenomena in nonequilibrium physics~\cite{strogatz2001exploring, de2023more, dorogovtsev2008critical}. For instance, they can realize Hopfield networks~\cite{hopfield1982neural} used for novel forms of computing, such as Ising machines~\cite{mohseni2022ising}.   
These networks also provide a testbed for studying collective phenomena such as synchronization~\cite{arenas2008synchronization}, cascade and critical transitions~\cite{scheffer2009early, scheffer2012anticipating, liu2024early, grziwotz2023anticipating,harris2024tracking, artime2024robustness}, and dynamical phases, e.g. Chimera states~\cite{abrams2004chimera,matheny2019exotic}.

Kerr parametric oscillators (KPOs) are a particularly promising building block for analog computation. A KPO can be realized by modulating the spring constant of a nonlinear resonator at approximately twice its resonance frequency~\cite{Lifshitz_Cross,bachtold2022mesoscopic,DykmanBook,eichler2023classical}. Above a certain modulation threshold, the system transitions into one of two stable oscillation `phase states', each with equal amplitude but opposite phase. These states can be mapped to the classical states of a spin--1/2 system (up and down), and can be used as binary logic elements~\cite{Mahboob_2008,Wilson_2010,Puri_2017,mahboob2016electromechanical,Nigg_2017,Nosan_2019,Frimmer_2019, Grimm_2019,wang_2019,Puri_2019_PRX,Miller_2019_phase,Yamaji_2023,Frattini_2024,alvarez2024biased,han2024coupled,margiani2025three} for a variety of applications~\cite{xue2025critical,szedlak2014control,mendoza2022intrinsic}. Additionally, these coupled bimodal and out-of-equilibrium resonators allow for studying complex dynamical phenomena in a highly controllable and stable platform.

\begin{figure}[!t]
	\centering
\includegraphics[width=0.85\columnwidth]{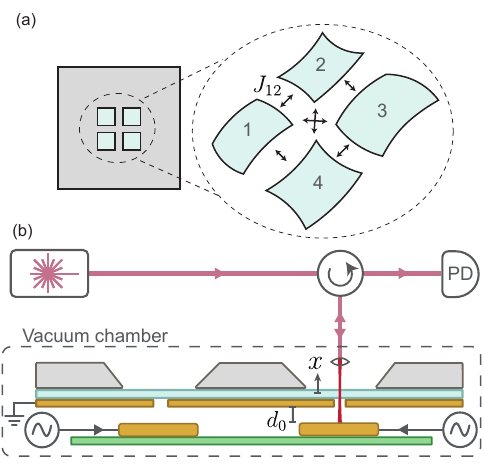}
	\caption{Coupled silicon nitride membranes. (a)~Four silicon nitride mechanical membranes (blue) are coupled through the silicon substrate (gray) to their neighbors with coupling strength $J_{ij}$. 
    (b)~The membranes are metallized and capacitively coupled to electrodes (gold) at a distance $d_0\approx \SI{13}{\micro\meter}$ below the membranes. We send 1550-nm laser light to a single membrane for interferometric readout. The laser is focused on the membrane with a GRIN lens, and the light reflected from the membrane is picked up by a circulator and collected by a photodiode (PD).} 
	\label{Fig:setup_diagram} 
\end{figure}

While early theory proposals~\cite{Goto_2016, Dykman_2018, Puri_2019_PRX} and demonstrations~\cite{Margiani_2023} are based on weakly coupled KPOs, much richer physics is found when the coupling strength is on the order of the damping rate~\cite{Heugel_2022,heugel2023proliferation,alvarez2024biased,ameye2025parametric,margiani2025three}. There, the interplay between nonlinearity and coupling gives rise to doubly unstable `ghost states'~\cite{heugel2023proliferation} and to additional states with mixed symmetry~\cite{Heugel_2022,margiani2025three}. Strong coupling is also essential for enabling coherent transitions between states in quantum annealing~\cite{Goto_2018}. Aside from strong coupling, detuning between the constituent oscillators leads to an asymmetric Ising model~\cite{han2024coupled}, which is central for modeling neural networks and other nonreciprocal systems~\cite{szedlak2014control,fruchart2021non}. Combining strong coupling and large \textit{in-situ} tuning in a scalable experimental platform would allow access to a range of important applications, including rapid phase logic operations~\cite{Nosan_2019,Frimmer_2019} and programmable neural networks~\cite{han2024coupled}. In reality, however, fulfilling all of these criteria in an experiment is difficult.

In this work, we present an electromechanical platform based on coupled silicon nitride membrane resonators on a single chip. Each resonator can be individually tuned in frequency by a few $\SI{}{\kilo\hertz}$, features a quality factor up to $Q \sim 10^4$, and exhibits strong  nearest-neighbor coupling, with the normal mode splitting comparable or larger than the membranes mechanical linewidth. The experimental platform is scalable, as many membranes can be fabricated on a single silicon substrate. We read out multiple mechanical resonators using a single laser interferometer, significantly reducing the experimental complexity. With this setup, we demonstrate a continuous crossover between a far detuned `resonator regime', with each mode comprising mostly one bare membrane oscillating, to a `normal-mode regime' where we observe strongly hybridized KPOs involving both membranes. The system can thus access various scenarios that have hitherto only been studied separately. In the future, this platform will allow us investigating many-body physics and computational architectures beyond the degenerate Ising system.

\textit{Device and readout.}--- Our device consists of a single silicon chip containing four low-stress silicon nitride membranes (Norcada NXVA-2250-5045A), see Fig.~\ref{Fig:setup_diagram}(a). The \SI{50}{\nano\meter}-thick membranes have lateral dimensions of $\SI{450}{\micro\meter} \times \SI{450}{\micro\meter}$ and are positioned \SI{50}{\micro\meter} apart from each other in a two-by-two grid. For electrostatic tuning and actuation, we coat the membranes with Ti (\SI{10}{\nano\meter}) and Au (\SI{30}{\nano\meter}) using optical lithography and e-beam evaporation, see Fig.~\ref{Fig:setup_diagram}(b) and Appendix~\ref{app:sec:fab}. The centers of the membranes, where we optically read out the membrane oscillations, are left uncoated to minimize heating due to light absorption. The chip is positioned and clamped on top of spacers on a printed circuit board (PCB), setting a distance $d_0\approx \SI{13}{\micro m}$ between the membrane surface and the PCB's electrodes. The membrane chip and PCB are placed in a vacuum chamber (pressure $\sim \SI{e-7}{\milli\bar}$) at room temperature.

\begin{figure}[t]
	\centering
	\includegraphics[width=1\columnwidth]{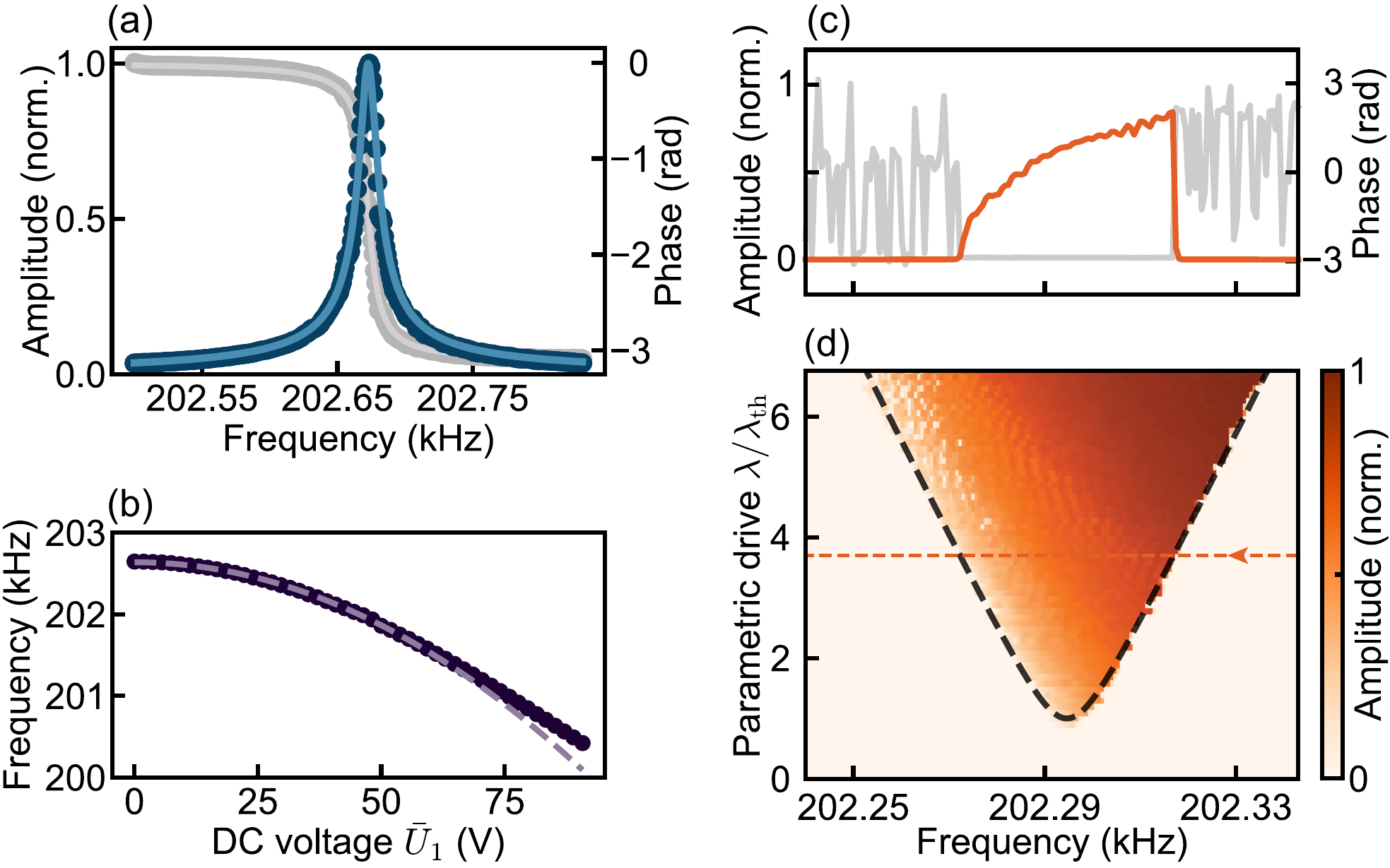}
	\caption{Characterization of single membrane resonator. (a)~Measured (dots) and fitted (solide lines) mechanical response amplitude (blue) and phase (grey) as a function of the driving frequency $\omega$. Fitting to the resonance curve $x_1(\omega) \propto (\omega_1^2 -\Omega^2 - i\omega\Gamma_1)^{-1} U_1(\omega)$ yields $\Omega_1/2\pi \approx \SI{202.67}{\kilo\hertz}$ and damping $\Gamma_1/2\pi = \SI{12.59(8)}{\hertz}$. (b)~Membrane frequency $\Omega_1+\delta\Omega_1$ as a function of applied DC voltage $\bar{U}_1$. From a quadratic fit, we extract the membrane-electrode distance $d_0 \approx \SI{13}{\micro\meter}$, see Appendix~\ref{app:sec:capacitive-driving}. (c)~Amplitude (orange) and phase (grey) measured at $\omega$ in response to parametric driving at $2\omega$ with $\lambda/\lambda_\mathrm{th} = 3.7$. Inside the large-amplitude lobe, the phase locks to one of the two phase states. (d)~Repeating frequency sweeps for multiple $\lambda/\lambda_\mathrm{th}$ reveals the characteristic `Arnold tongue'. The sweep in (c) is marked by a dashed orange line. A black dashed line marks the boundary of the Arnold tongue, using the fitted value from the linear sweep. }
	\label{Fig:lin_1weep_tongue} 
\end{figure}

We read out the motion of a single membrane with an interferometer. Laser light (Rio Orion \SI{1550}{nm}) with a wavelength of \SI{1550}{\nano\meter} is sent through a fiber to a circulator (Koheron FOCIR1550-A) and into the vacuum chamber with a fiber feedthrough, and couples to free space via an optical fiber end with a 50\% reflection coating (Thorlabs, custom made). The 50\% transmitted light is focused by a gradient refractive index lens (Thorlabs GRIN2915) onto the uncoated center of the membrane. We measure the total reflected light with a photodiode (PD, Koheron PD100B) and a lock-in amplifier (Zurich Instruments MFLI). The interference between the light reflected from the membrane and from the coated fiber provides a record of the membrane displacement $x$.

\textit{Single membrane resonator.}--- We first characterize the linear and parametric responses of a single membrane. The other membranes being far detuned in frequency, we observe the motion of the mode consisting mostly of the membrane we read out. We consider the displacement $x_i$ of the fundamental out-of-plane vibration mode of membrane $i$, modeled by the resonator equation of motion 
\begin{align}
\ddot{x}_i+\Omega_i^2\left[1- \lambda_i \cos \left(2 \omega t\right )\right] &x_i+\Gamma_i \dot{x}_i+\beta_i x_i^3 \nonumber \\
&+ \sum_{j\neq i} J_{ij}^2x_j=F_i(t)/m_i,
\label{eq:EOM}
\end{align}
where dots indicate derivation with respect to time $t$, $\Omega_i$ is the resonance frequency, $\Gamma_i$ the damping rate,  $\beta_i$ the geometrical Duffing nonlinearity, and $J_{ij}=J_{ji}$ the coupling rate to membrane $j$. The coupling $J_{ij}$ can be tuned by varying the distance between the membranes.  The resonator is driven by a near-resonant force
$F_i = \tilde{F}_i\cos(\omega t)$ with $\omega\approx\Omega_i$, and a parametric drive with modulation depth $\lambda_i$. The effective mass $m_i\approx \SI{38}{\nano\gram}$ of each membrane is estimated from the thicknesses of the silicon nitride and the metal layers, see Appendix~\ref{app:sec:effective-mass}.

We measure the linear response of membrane 1 by applying a voltage $U_1 = \bar{U}_1+ \tilde{U}_{\omega,1} \cos ( \omega  t)$ to the electrode underneath the membrane, with $\bar{U}_1$ a DC voltage, and $F_1 \propto  \tilde{U}_{\omega, 1}\cos(\omega t)$. We measure the complex mechanical response $x_1(\omega)$ while sweeping $\omega$, see Fig.~\ref{Fig:lin_1weep_tongue}(a). For small enough $\tilde{U}_{\omega, {1}}$ and $\lambda_1 = 0$, the response is small and we can approximate $\beta_1 x_1^3 = 0$. From a fit to the measured response, we extract a fundamental mode frequency $\Omega_1/2\pi = \SI{202.67}{\kilo\hertz}$ and decay rate $\Gamma_1/2\pi = \SI{12.59(8)}{\hertz}$, yielding a quality factor $Q_1 \equiv \Omega_1/\Gamma_1= \SI{16.1(1)e3}{}$. For these measurements, all the other membranes are far detuned in frequency, so there is barely any mode hybridization. In a similar fashion, we characterize membrane 2, finding a resonance frequency $\Omega_2/2\pi \approx \SI{203.40}{\kilo\hertz}$, a decay rate $\Gamma_2/2\pi = \SI{39.2(6)}{\hertz}$, and a quality factor $Q_2 =\SI{5.19(8)e3}{}$, see Appendix~\ref{app:sec:charac2-3}. We normalize all measured amplitude due to the difficulty of calibrating from voltage to position in our apparatus, see Appendix~\ref{app:sec:double-interferometer}.

The voltage $U_i$  can also be used to tune each membrane frequency $\Omega_i$~\cite{zhou2021high,schmid2014single, Bagci2014Optical,puglia2025room}. This voltage generates a capacitive force $F_{\mathrm{c},i}$ which leads to an electrical spring constant $k_{\mathrm{c},i} = - \partial F_{\mathrm{c},i}/\partial x_i=  U_i^2\partial^2_{x_i} C_i/2 $, where the electrode-membrane capacitance $C_i$ is inversely proportional to their separation, $C_i\propto (d_0 + x_i)^{-1}$, see Appendix~\ref{app:sec:capacitive-driving}. This shifts $\Omega_i = \sqrt{k_i/m_i}$ by 
\begin{align}\label{eq:delta_omega}
    \delta \Omega_i = \sqrt{\frac{k_i + k_{\mathrm{c},i}}{m_i}} - \sqrt{\frac{k_i}{m_i}}\approx \frac{1}{4m_i\Omega_i}  \frac{\partial^2C_i}{\partial x_i^2}U_i^2 ,
\end{align}
in the limit $k_{\mathrm{c},i}\ll k_i$, with $k_i$ the bare mechanical spring constant of the membrane. In Fig.~\ref{Fig:lin_1weep_tongue}(b), we show $\Omega_1$ being tuned by a DC voltage $\bar{U}_1$. Up to $\SI{50}{\volt}$, we observe the quadratic dependency $\delta \Omega_1\propto 
\bar{U}_1^2$ expected from Eq.~\eqref{eq:delta_omega}, with a tunability of $\delta \Omega_1 >\SI{2}{\kilo\hertz}$. At higher $\bar{U}_1$, the frequency shift does not follow Eq.~\eqref{eq:delta_omega} precisely. We suspect this stiffening to be caused by the positive Duffing nonlinearity at large equilibrium positions, see Appendix~\ref{app:sec:capacitive-driving}.

We drive the membrane parametrically by applying both a DC voltage $\bar{U}_i$ and a time-varying (AC) voltage at frequency $2\omega$ with amplitude $\tilde{U}_{2\omega, {i}}$, while we set $F_1=0$  (i.e., there is no voltage at frequency $\omega$). With a total voltage  $U_i(t) = \bar{U}_i+ \tilde{U}_{2\omega, {i}}\cos(2\omega t)$ and using  Eq.~\eqref{eq:delta_omega} for $\bar{U}_i \gg \tilde{U}_{2\omega, {i}}$, we  approximate the induced frequency shift as
\begin{align}\label{eq:delta_omega_U_Exp}
    \delta \Omega_i (t) \approx  \frac{1}{4m_i\Omega_i}  \frac{\partial^2C_i}{\partial x_i^2} \left [\bar{U}^2_i+ \bar{U}_i \tilde{U}_{2\omega, {i}}\cos(2\omega t) \right ] ,
\end{align}
with the second term corresponding to an effective $\lambda\propto \tilde{U}_{2\omega, {i}}$ in Eq.~\eqref{eq:EOM}. We record the membrane's oscillation amplitude and phase while slowly sweeping $\omega$ from high to low frequencies, i.e., in the direction opposed to the Duffing nonlinearity, see Fig.~\ref{Fig:lin_1weep_tongue}(c). 
Within a frequency window around $\Omega_1$, we observe parametric oscillations with large amplitude (limited by $\beta_1$) and a phase locked to one of the two phase states (differing  by phase $\pi$)~\cite{eichler2023classical}. Outside the window, the response amplitude is zero and the phase is undefined. Repeating parametric sweeps for various values of $\tilde{U}_{2\omega, 1}$, we obtain an `Arnold tongue', see Fig.~\ref{Fig:lin_1weep_tongue}(d).  We label the vertical axis by the normalized quantity $\lambda/\lambda_\mathrm{th}$, where $\lambda_\mathrm{th}$ is the threshold value above which parametric oscillations are measured within a frequency sweep. This phase diagram indicates the region where the response is finite and the KPO is in one of the two phase states. 

\begin{figure}[t]
	\centering
	\includegraphics[width=0.9\columnwidth]{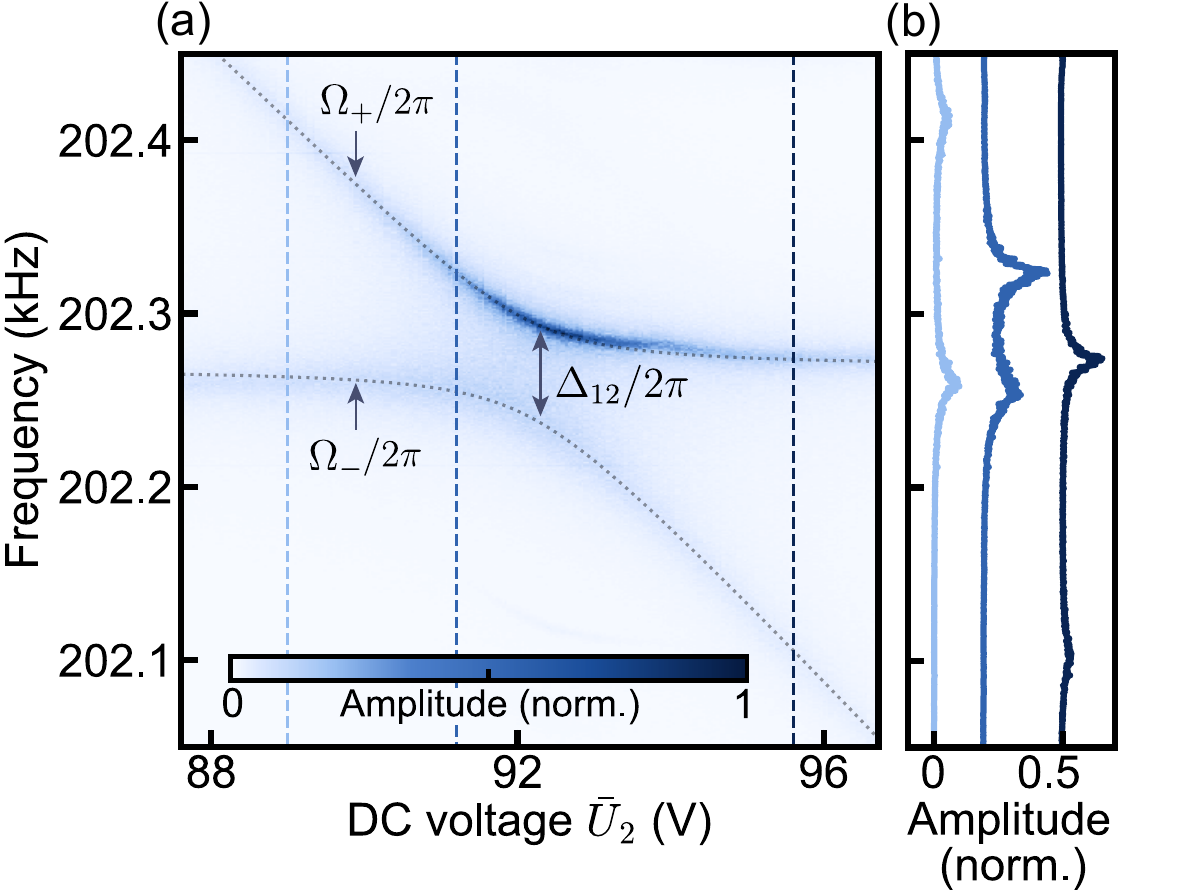}
	\caption{Avoided crossing between the fundamental modes of membranes 1 and 2. (a)~Displacement amplitude spectral density (ASD) of  membrane 1 as a function of frequency and $\bar{U}_2$ applied to membrane 2, with fixed $\bar{U}_1 = \SI{34.1}{\volt}$. Dotted lines indicate fits of $\Omega_\pm/2\pi$, yielding an  avoided crossing  gap $\Delta_{12}/2\pi = \SI{55.2(5)}{\hertz}$, cf. Eq.~\eqref{eq:omega_-+}. (b)~Displacement ASD line cuts for different values of $\bar{U}_2$, corresponding to the vertical dashed lines in panel (a). The amplitudes are offset for better visibility.     }
	\label{Fig:avoided_crossing} 
\end{figure}

\textit{Coupled membranes.}--- We now probe the response of two neighboring membranes (1 and 2), coupled through their substrate~\cite{de2022mechanical,correia2024coherent, madiot2021bichromatic} and tuned by separate DC voltages $\bar{U}_{1,2}$. From Eq.~\eqref{eq:EOM}, we find the system's eigenfrequencies, see Appendix~\ref{app:sec:avoided-crossing},
\begin{align}\label{eq:omega_-+}
    \Omega_\pm = \left[ \frac{\Omega_1^2 +\Omega_2^2}{2} \pm \frac{\sqrt{ (\Omega_1^2-\Omega_2^2)^2 + 4J_{12}^4}}{2}\right]^{1/2},
\end{align}
which, for $\Omega_1 \approx \Omega_2$ and in the limit $J_{12}\ll \Omega_1$, simplifies to $ \Omega_\pm  =  \Omega_1 \mp J_{12}^2 / (2\Omega_1)$. This yields a characteristic avoided crossing with a gap $\Delta_{12} \equiv \Omega_- - \Omega_+ = J_{12}^2/\Omega_1$. The $+$ ($-$) mode corresponds to the symmetric (anti-symmetric) mode $x_1 + x_2$ ($x_1 - x_2)$. To measure this gap, we keep $\bar{U}_1 = \SI{34.1}{\volt}$ constant and measure the spectral response of membrane 1 to a white force noise for various values of $\bar{U}_2$, see Fig.~\ref{Fig:avoided_crossing}.

Fitting Eq.~\eqref{eq:omega_-+} to our data, we obtain a coupling {$J_{12}/(2\pi) = \SI{3.34(3)}{\kilo\hertz}$ and splitting $\Delta_{12}/2\pi = J_{12}^2/(2\pi\Omega_1)= \SI{55.2(5)}{\hertz}$. This splitting is on the order of the combined membranes' mechanical linewidth $(\Gamma_1 + \Gamma_2)/2\pi = \SI{51.8(7)}{\hertz}$. With an analogous measurement, we  characterize the coupling between membrane 1 and its diagonally opposed membrane 3 to be $J_{13}/(2\pi) = \SI{2.77(3)}{\kilo\hertz}$, with normal mode splitting $\Delta_{13}/2\pi=\SI{37.9 \pm 0.4}{\Hz}$, see Appendix~\ref{app:sec:coupling1-3}. As expected, this coupling is smaller due to the larger separation between those two membranes.

\begin{figure*}[t]
	\centering
	\includegraphics[width=0.75\textwidth]{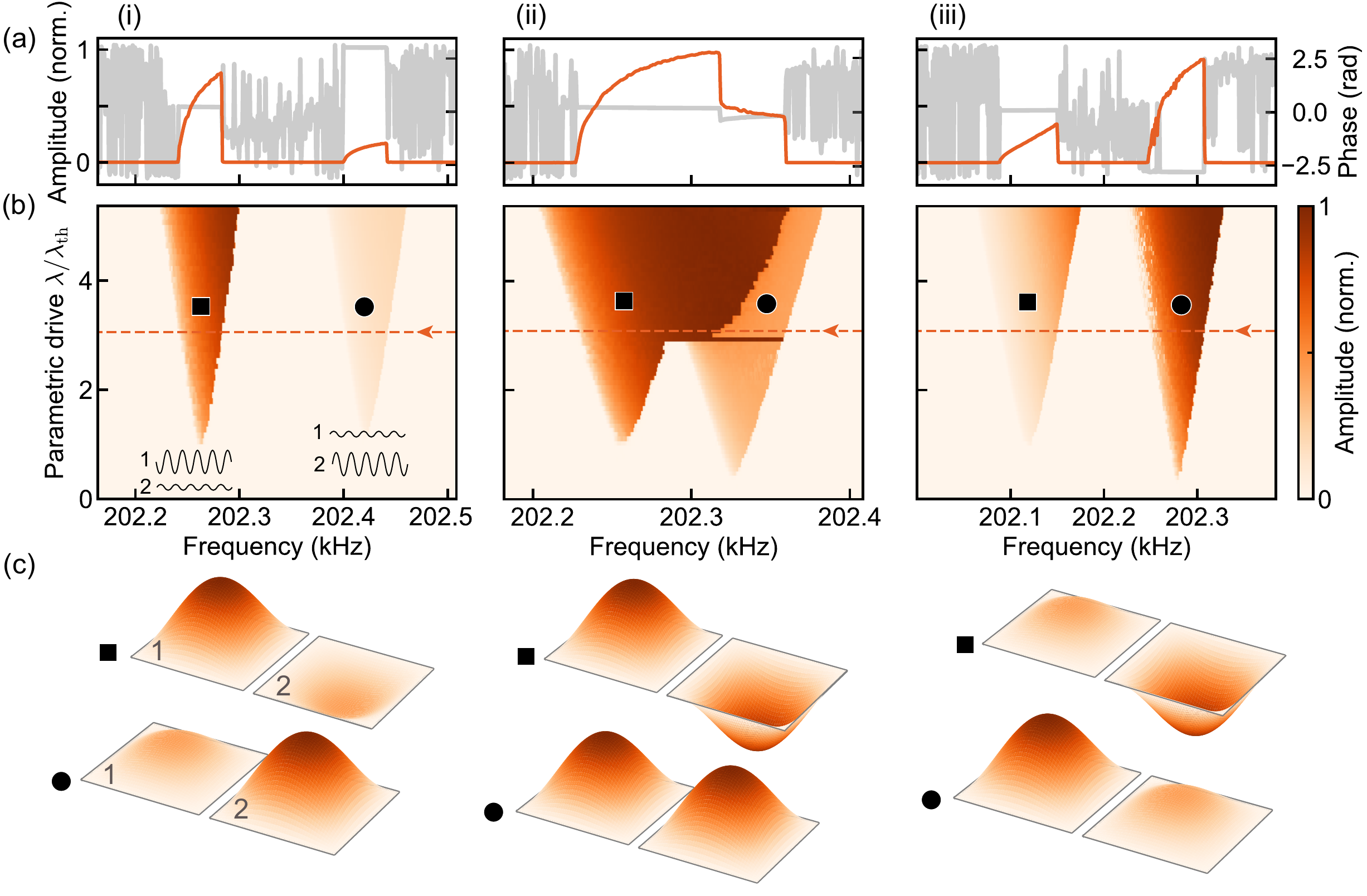}
	\caption{Parametric response of two coupled membranes (1 and 2). Columns (i)-(iii) correspond to three different values of $\bar{U}_2$, cf. vertical lines in Fig.~\ref{Fig:avoided_crossing}. (a)~Amplitude (orange) and phase (gray) measured on membrane 1 with $\lambda/\lambda_\mathrm{th} = 3.1$ and $\bar{U}_1=\SI{34.1}{\volt}$. The frequency was swept from high to low values. (b)~Response to parametric driving of the coupled membranes for various $\lambda/\lambda_\mathrm{th}$. The data in (a) was taken along the dashed orange lines. An inset in panel (i) illustrates $x_1$ and $x_2$ as a function of time.  (c)~Qualitative mode shape of each membrane when driven at the positions in (b) indicated by the square and circle shapes. }
	\label{Fig:multiple_tongues} 
\end{figure*}

So far, we have shown that our system of membrane resonators has mechanical modes with high quality factors whose frequencies can be widely tuned and whose coupling is on the order of the total damping rate. We have also seen that these modes individually respond to parametric driving. Now, we combine all of these elements, demonstrating a network of KPOs that can be continuously tuned between two very different regimes: in the first regime, the original membrane resonators (1 and 2) are detuned, and the system response can be well understood in terms of parametric oscillations localized on either membrane. In the second regime, the resonators form hybridized modes and the parametric response involves both membranes simultaneously, generating a rich phase diagram of overlapping Arnold tongues.

In Fig.~\ref{Fig:multiple_tongues}, we parametrically drive the two membranes simultaneously with strength $\lambda$ and frequency $2\omega$ while measuring the amplitude and phase of membrane 1 at $\omega$. For each measurement, we set $F_i=0$ and apply fixed DC voltages $\bar{U}_1$ and $\bar{U}_2$ to the two membranes. We select three different values for $\bar{U}_2$, corresponding to three different points along the avoided crossing, see vertical lines in Fig.~\ref{Fig:avoided_crossing}(a). Sweeping $\omega$ from higher to lower frequencies, we obtain Fig.~\ref{Fig:multiple_tongues}(a). Repeating the parametric drive sweeps for multiple values of $\lambda$, we recover the colorplots in Fig.~\ref{Fig:multiple_tongues}(b). Note that the AC voltage amplitudes sent to each membrane needs to be tuned to compensate for mode hybridization and difference in quality factors, see Appendix~\ref{app:sec:param_matching}.

In Figs.~\ref{Fig:multiple_tongues}(i) and (iii), we measure two tongue-shaped regions with finite amplitudes and defined phases. The tongues are centered around the two mode frequencies measured for the corresponding voltages in Fig.~\ref{Fig:avoided_crossing}, with the tongue around $\Omega_1$ exhibiting a stronger signal than that at $\Omega_2$. We interpret these results the following way: here, the two membranes are detuned from each other, and we predominantly drive only one of them at each frequency. This produces large oscillations of membrane 1 around \SI{202.26}{\kilo\hertz} in Fig.~\ref{Fig:multiple_tongues}(i) and around \SI{202.28}{\kilo\hertz} in Fig.~\ref{Fig:multiple_tongues}(iii). Weak oscillations of membrane 1 can still be seen in the other tongue around \SI{202.41}{\kilo\hertz} in Fig.~\ref{Fig:multiple_tongues}(i) and around \SI{202.12}{\kilo\hertz} in Fig.~\ref{Fig:multiple_tongues}(iii). This is due to the off-resonant force it experiences through the coupling term $J_{12}^2x_2$ in the presence of parametric oscillation of membrane 2,  cf. Eq.~\eqref{eq:EOM}. Inset in Fig.~\ref{Fig:multiple_tongues}(i)(b) illustrate $x_1$ and $x_2$ as a function of time for the different lobes, while visualizations of the spatial mode shapes are shown in Fig.~\ref{Fig:multiple_tongues}(c).

When the membrane resonators are tuned into an avoided crossing, they become hybridized. Parametrically driving these normal modes results in a pattern of overlapping Arnold tongues~\cite{Heugel_2022}. The Arnold tongues with lower and higher center frequency represent antisymmetric ($x_1-x_2$) and symmetric ($x_1+x_2$) parametric oscillation states, respectively, see Appendix~\ref{app:sec:avoided-crossing}. Transitions between states can be measured as jumps in the oscillation amplitude. For the line sweep in Fig.~\ref{Fig:multiple_tongues}(a)(ii), a jump to the symmetric state occurs at approximately \SI{202.35}{\kilo\hertz}, below which the zero-amplitude state becomes unstable and parametric oscillations start. A second jump is observed at
\SI{202.32}{\kilo\hertz}, well within the symmetric Arnold tongue. 

To the left of this jump, the interplay between the nonlinearity and the resonator coupling destabilizes the symmetric state, forcing the system to jump to the antisymmetric state~\cite{Heugel_2022,ameye2025parametric}. Those intricate patterns were previously only measured with electrical resonators whose quality factors were much lower than those of the mechanical modes reported here~\cite{Heugel_2022,alvarez2024biased,margiani2025three}. Confirming their existence in a high-$Q$ system is a first step towards studying many-body fluctuations~\cite{heugel2023role} and interstate statistical transitions~\cite{heugel2023proliferation} in nanomechanical systems. Such dynamics can be measured with a lock-in amplifier when a mode's ringdown time ($2Q/\Omega_i$) is longer than the required integration time $\tau$. With electrical systems, which typically have ringdown times on the order of \SI{25}{\micro\second}~\cite{Frimmer_2019}, this requirement is harder to fulfill than in the present system, which features $\tau_1=2Q_1/\Omega_1\approx\SI{26}{\milli\second}$ and $\tau_2=2Q_2/\Omega_2\approx\SI{8}{\milli\second}$.

We emphasize that even though we only measure the amplitude and phase of membrane 1, we gain access to much information about the entire coupled system. In the detuned case in Fig.~\ref{Fig:multiple_tongues}(i) and (iii), the coupling allows us to indirectly measure the parametric oscillation of membrane 2 via the motion of membrane 1. In Fig.~\ref{Fig:multiple_tongues}(ii), we can detect transitions between different parametric oscillation states of the network via jumps in the amplitude. Together with analytical and numerical modeling~\cite{Heugel_2022,ameye2025parametric}, such data enables us to understand the underlying phase diagram of the KPO network, i.e., which oscillation states exist where in the space spanned by $\omega$ and $\lambda$. The amplitude can then be used as a proxy measurement to differentiate between different states, for example in the context of KPO-based Ising simulators~\cite{margiani2025three}. To calibrate and directly confirm such experiments, our setup could be combined with wide-field stroboscopic imaging to measure the displacement of all the membranes simultaneously~\cite{conway2007stroboscopic}.

\textit{Summary and outlook.---}  We demonstrated a system of coupled nonlinear mechanical resonators based on silicon nitride membranes. Coupling between membranes can be designed via their separation. The platform combines high quality factors and large individual tuning of the resonator frequencies, offering access to distinct regimes of dynamics. In addition, we show that information about the entire network can be inferred from a measurement of just one membrane.

This platform is easily scalable. We envision that networks of six or nine membrane resonators can be probed without significant additional technical overhead, assuming that existing theory models~\cite{ameye2025parametric} appropriately describe the emerging phenomena. Beside the number of membranes, the quality factors of individual membranes can potentially be increased by orders of magnitudes using soft clamping~\cite{Tsaturyan2017Ultracoherent,mashaal2025strong}. This will greatly facilitate the study of activated fluctuations between the different states in such complex systems~\cite{heugel2023proliferation}. When detuning the constituent KPOs, the system could potentially mimic an asymmetric Ising model, which is central to the understanding of neural networks~\cite{han2024coupled}. Furthermore, we plan to use this platform as a testbed for Ising machines~\cite{mohseni2022ising}, Hopfield networks~\cite{hopfield1982neural}, and the study of emergent nonlinear collective phenomena~\cite{arenas2008synchronization, scheffer2009early, scheffer2012anticipating, liu2024early, grziwotz2023anticipating, harris2024tracking, artime2024robustness}, dynamical phases~\cite{abrams2004chimera, matheny2019exotic}, and topology in nonlinear networks~\cite{smirnova2020nonlinear, ravets2025thouless, villa2024topological}. With this, nanomechanical KPO networks will become a useful tool for tackling open questions in complex systems and artificial neural networks.

\textit{Acknowledgments.---} We thank R. Pachlatko, L. Ziegler, and N. Prumbaum for insightful discussions. V.D. acknowledges support from the ETH Zurich Postdoctoral Fellowship Grant No. 23-1 FEL-023 and the Swiss National Science Foundation (SNSF) Postdoctoral Fellowship Grant 217118. A.E. acknowledges financial support from the Swiss National Science Foundation (SNSF) through the Sinergia Grant No.~CRSII5\_206008/1.

\appendix

\section{Membrane metallization}\label{app:sec:fab}
We metallize the silicon nitride membranes for capacitive actuation, see Fig.~\ref{app:Fig:PCB_photo}(a). All the membranes are coated with titanium (for adhesion) and gold (for electrical conduction), and connected to a common ground. We leave the center of each membrane uncoated in order to reduced laser absorption which we found to lead to large mechanical frequency shifts (kilohertz-scale mechanical resonance shifts for milliwatt-level laser power). The electrodes on the PCB allow to individually address each membrane. 

To metallize the membranes, we first attach the \SI{5}{\milli\meter}$\times$\SI{5}{\milli\meter} silicon chip to a silicon wafer with Crystal bond. We spincoat  \SI{450}{\nano\meter} of photosensitive liftoff resist (LOR5B) for \SI{40}{\second} at \SI{4000}{rpm}, followed by a \SI{300}{\second} bake-out on a hot plate at \SI{170}{\celsius}. We then spincoat a \SI{1.8}{\micro\meter} thick layer of the photoresist AZ1518 (\SI{40}{\second} spinning at \SI{4000}{rpm} and bake-out for \SI{90}{\second} at \SI{100}{\celsius}). We define a pattern by laser writing (Heidelberg DWL66+) with a laser wavelength \SI{375}{\nano\meter}, write speed \SI{110}{\milli\meter^2/min}, and  laser power \SI{55}{\micro\watt}. We then develop the two resists in AZ400k for \SI{2}{mins}. We transfer the sample to deionized water, and then dry it with a nitrogen gun. We further clean leftover resist residues in an O$_2$ asher for \SI{120}{\second} at \SI{100}{\watt}. After this process, we detach the membrane from the silicon holder by placing it on a hotplate at \SI{70}{\celsius}. We use e-beam evaporation to deposit \SI{10}{\nano\meter} of Ti followed by \SI{30}{\nano\meter} of Au. We lift off the remaining resist by placing the device in a dimethyl sulfoxide (DMSO) bath at \SI{80}{\celsius} for \SI{30}{mins}. Finally, we rinse and clean the device with acetone, isopropyl alcohol (IPA) and deionized (DI) water, and blow it dry with a nitrogen gun. The membrane chip is then clamped to the PCB, see Fig.~\ref{app:Fig:PCB_photo}(b).

\begin{figure}[t]
	\centering
	\includegraphics[width=1\columnwidth]{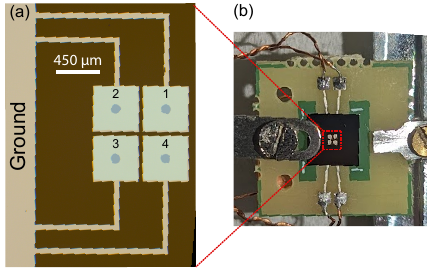}
	\caption{Parametric membranes device. (a)~Microscope image of the metallized membranes with labels 1-4. All membranes share a common ground and have their center uncoated. (b)~Photograph of the membrane chip after it is clamped to the PCB.} 
	\label{app:Fig:PCB_photo} 
\end{figure}

\section{Membrane effective mass}\label{app:sec:effective-mass}
Each membrane consists of a 50-nm-thick silicon nitride film (density \SI{3170}{\kilo\gram/\meter^{3}}), coated with a 10 nm thick titanium layer (density \SI{4506}{\kilo\gram/\meter^{3}}) and a 30 nm thick gold layer (density \SI{19300}{\kilo\gram/\meter^{3}})~\cite{density_Ti_Au}. The silicon nitride has an area of $(\SI{0.45}{\milli\meter})^2 = \SI{0.20}{\milli\meter^2}$, while the metal coatings cover a slightly smaller area of \SI{0.19}{\milli\meter^2} due to the central hole with a radius of \SI{50}{\micro\meter}. Summing the masses of these layers gives a total mass of approximately \SI{150}{\nano\gram}. The effective mass of the fundamental mode is approximately a quarter of this total mass~\cite{hauer2013general}, i.e., $m \approx \SI{38}{\nano\gram}$. 

\section{Capacitive driving}\label{app:sec:capacitive-driving}
Here, we calculate the capacitive force $F_{\mathrm{c},i}$  and the effective spring constant $k_{\mathrm{c},i}$ induced on  membrane $i$ by applying a voltage $U_i$ on the electrode beneath it.

\textbf{Capacitance:} We consider the capacitance between two conducting plates of area $A = (\SI{450}{\micro\meter})^2$, i.e. the membrane and electrode metallization area (neglecting the metallization hole of radius $\SI{50}{\micro\meter}$), which are separated by a distance $d_i$. This leads to capacitance~
\begin{align}
    C_i = \epsilon_0 \epsilon_\mathrm{r} \frac{A}{d_i},
\end{align}
 with $\epsilon_0 = 8.85\times 10^{-12}$ F/m the vacuum permittivity, and $\epsilon_\mathrm{r} =1$ the relative permittivity of the surrounding medium (vacuum).

 We need to take into account the fundamental membrane mode-shape $u(y,z)$ in the $y$-$z$ membrane plane, which is~\cite{schmid2016fundamentals}
\begin{align}
    u(y,z) = \sin \left(\frac{\pi y}{L} \right) \sin \left(\frac{\pi z}{L} \right),
\end{align}
assuming clamped boundary conditions for the membrane and with $L=\SI{450}{\micro\meter}$  the membrane lateral dimension. In this paper, we only consider the fundamental mode of each membrane, but a similar analysis can be done for higher order mechanical modes. The membrane-electrode separation is $d_i(y,z) \equiv  d_0 + u(y,z) x_i$, with $d_0$ the distance between the membrane and the electrode, and $x_i$ the membrane displacement at its center. The capacitance is then~\cite{puglia2025room}
\begin{align}
    C_i =  \int_0^L\int_0^L  \frac{\epsilon_0 \epsilon_\mathrm{r}}{d_0 + u(y,z)x_i} dydz.
\end{align}
\textbf{Capacitive force:} The potential energy of the capacitor is $ U_{\mathrm{pot},i}= -C_i U_i^2/2$ for a voltage $U_i$ applied to the membrane's electrode, yielding an attractive force 
\begin{align}
    F_{\mathrm{c},i}(x_i) &= -\frac{\partial U_{\mathrm{pot},i}}{dx_i} = \frac{1}{2}  \frac{\partial C_i}{\partial x_i} U_i^2   \\
    &= - \frac{1}{2} U_i^2 \int_0^L\int_0^L  \frac{\epsilon_0 \epsilon_\mathrm{r}u(y,z)}{[d_0 + u(y,z)x_i]^2} dydz.
\end{align}
\textbf{Spring constant:} At the equilibrium position $x_\mathrm{eq}$, this force can be approximated as
\begin{align}\label{app:eq:F_taylor}
     F_{\mathrm{c},i}(x_i)  \approx   F_{\mathrm{c},i}(x_\mathrm{eq}) + (x_i-x_\mathrm{eq}) \left. \frac{ \partial F_{\mathrm{c},i}}{\partial x_i}\right |_{x_i=x_\mathrm{eq}},
\end{align}
for $x_i \ll d_0$. When inserted in the equations of motion, the second term leads to a capacitive spring constant
\begin{align}
     k_{\mathrm{c},i} &= -  \left. \frac{ \partial F_{\mathrm{c},i}}{\partial x_i}\right |_{x_i=x_\mathrm{eq}}  \\
    &=  -U_i^2 \int_0^L\int_0^L  \frac{\epsilon_0 \epsilon_\mathrm{r}u^2(y,z)}{[d_0 + u(y,z)x_\mathrm{eq}]^3} dydz,
\end{align}
generating a spring softening. In the limit of small changes in equilibrium position, $x_\mathrm{eq}\ll d_0$, the spring constant becomes
\begin{align}
    k_{\mathrm{c},i} 
    &\approx  -\frac{\epsilon_0 \epsilon_\mathrm{r}}{d_0^3} U_i^2 \int_0^L\int_0^L  u^2(y,z) dydz\\
    &=  -\frac{\epsilon_0 \epsilon_\mathrm{r}A}{4d_0^3} U_i^2.
\end{align}

\textbf{Frequency shift:} As mentioned in the main text, $k_{\mathrm{c},i}$ is an additional spring constant acting on the mechanical resonator, leading to an effective spring constant $k_{i} + k_{\mathrm{c},i}$ and a shifted mechanical frequency $\Omega_{\mathrm{eff},i}=\sqrt{(k_{i} + k_{\mathrm{c},i})/m_i}$. The mechanical frequency shift is 
\begin{align}
    \delta \Omega_i = \Omega_{\mathrm{eff},i} - \Omega_i =  \sqrt{\frac{k_i + k_{\mathrm{c},i}}{m_i}} - \sqrt{\frac{k_i}{m_i}},
\end{align}
which, in the limit of $k_{\mathrm{c},i}\ll k_{i}$, reduces to
\begin{align}\label{app:eq:delta_omega}
    \delta \Omega_i     &\approx \frac{k_{\mathrm{c},i}}{2m_i\Omega_i}    =- \frac{\epsilon_0 \epsilon_\mathrm{r} A }{8m_i\Omega_i d_0^3}     U_i^2.
\end{align}
Thus, the frequency of the resonator decreases quadratically with voltage $ U_i$ applied to the electrode underneath it.

\textbf{Membrane-electrode distance:} Equation~\eqref{app:eq:delta_omega} provides a way of extracting the membrane-electrode distance $d_0$. Using the membrane area $A = \SI{0.2}{mm^2}$, the effective mass $m_i \approx \SI{38}{\nano\gram}$ derived in Appendix~\ref{app:sec:effective-mass}, and the prefactor $B \equiv - \epsilon_0 \epsilon_\mathrm{r} A/ (8m_i\Omega_i d_0^3) =2 \pi\times\SI{ 0.309(1)}{\hertz/\volt^2} $ from the fit in Fig.~\ref{Fig:lin_1weep_tongue}(b) relating the change in mechanical frequency for an applied voltage $\bar{U}_i$, we obtain $d_0 \approx \SI{13}{\micro\meter}$. At that distance, the membrane-electrode capacitance  is $C_i \approx \SI{140}{\femto\farad}$.

\textbf{Equilibrium position:} We note that the first term, i.e.  $F_{\mathrm{c},i}(0)$, in eq.~\eqref{app:eq:F_taylor} leads to a shift of the membrane's equilibrium position to $x_{\mathrm{eq},i}$, which can cause an increased mechanical frequency (stiffening) due to the positive Duffing nonlinearity $\beta x_i^3$. To see this, we first find the equilibrium position $x_{\mathrm{eq},i}$ where $F_{\mathrm{c},i}(0) -m_i\Omega_i^2=0$,
 \begin{align}
     x_\mathrm{eq} =-\frac{1}{2m_i\Omega_i^2} \frac{\partial C_i}{\partial x_i} U_i^2 \approx -\frac{2\epsilon_0 \epsilon_\mathrm{r}A}{\pi^2m_i\Omega_i^2 d_0^2} U_i^2
 \end{align}
in the limit $x_i \ll d_0$, and using $\partial C_i/\partial x_i = 4\epsilon_0 \epsilon_r A/(\pi^2 d_0^2)$.

We can redefine $x_i$ around this equilibrium position $x_i \rightarrow x_{\mathrm{eq},i} + x_i$, the Duffing term $\beta x_i^3$ then becomes 
\begin{align}
    \beta (x_{\mathrm{eq},i} + x_i)^3 =  \beta x_{\mathrm{eq},i}^3  + 3\beta x_{\mathrm{eq},i}^2 x_i + 3 \beta x_{\mathrm{eq},i} x_i^2  + \beta  x_i^3 .
\end{align}
The first term $\beta x_{\mathrm{eq},i}^3$ leads to a shift in equilibrium position, the second term $3\beta x_{\mathrm{eq},i}^2 x_i $ to a spring constant shift $3\beta x_{\mathrm{eq},i}^2$, the third term  $3 \beta x_{\mathrm{eq},i} x_i^2$ to a static force and dynamics at $2\omega$ for $x_i\sim\cos(\omega t)$, which gets filtered out by measuring with our lock-in at frequency $\omega$~\cite{eichler2013symmetry} and  effectively leads to a renormalization of the Duffing nonlinearity~\cite{kozinsky2006tuning,lifshitz2008nonlinear}, and the fourth term $3 \beta x_i^3$ to the usual Duffing nonlinearity.

\begin{figure}[t]
	\centering
    \includegraphics[width=0.7\columnwidth]{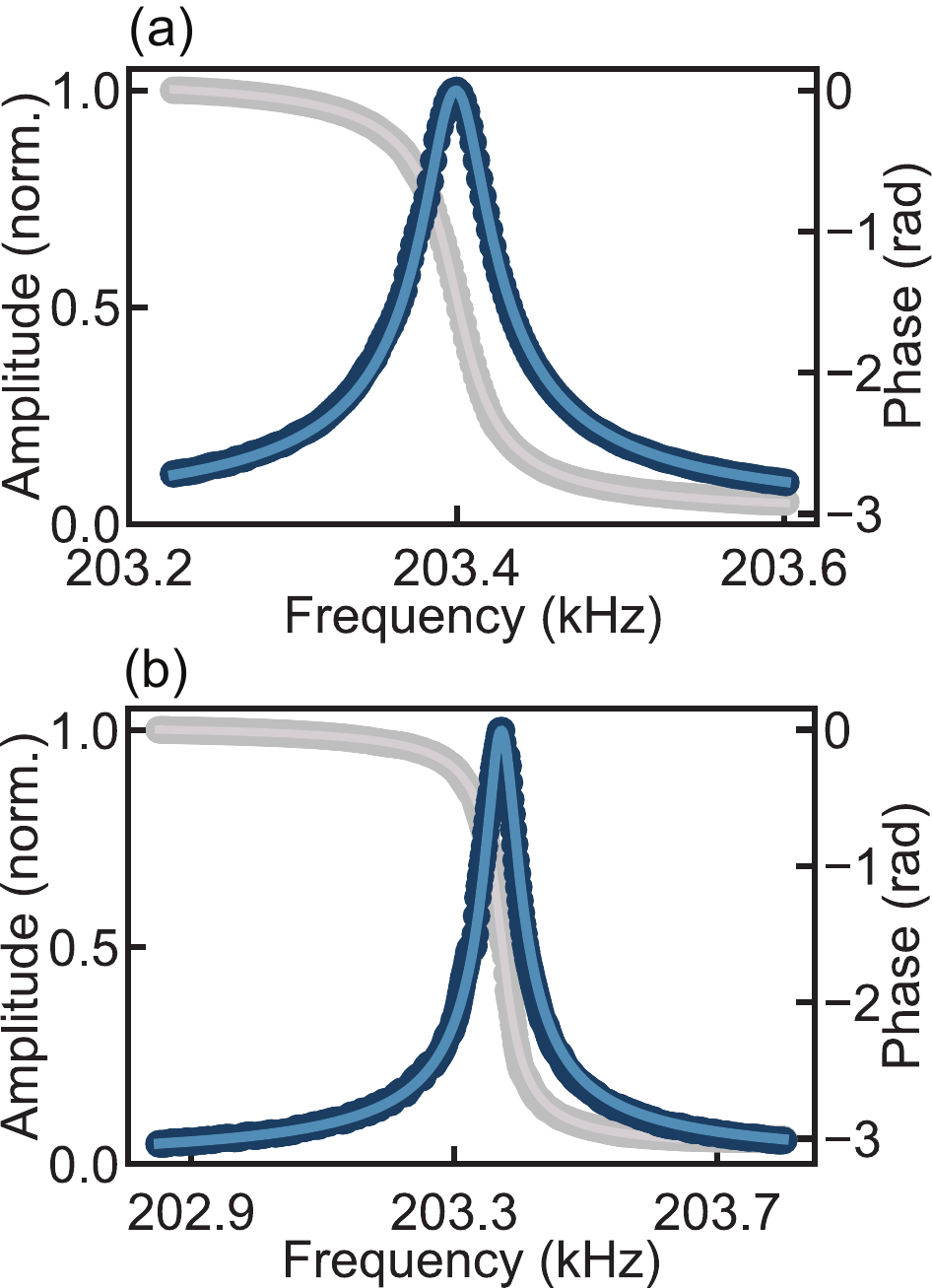}
	\caption{Linear mechanical response of membranes 2 and 3. Measured (dots) and fitted (solid lines) mechanical response amplitude (blue) and phase (grey) as a function of the driving frequency $\omega$ for (a)~membrane 2 and (b)~membrane 3. From the fits, we extract the frequency $\Omega_2/2\pi\approx\SI{203.40}{\kilo\hertz}$  and damping rate $\Gamma_2/2\pi = \SI{39.2(6)}{\hertz}$ (quality factor $Q_2=\SI{5.19(8)e3}{}$) of membrane 2, and the frequency $\Omega_3/2\pi\approx\SI{203.37}{\kilo\hertz}$ and damping rate $\Gamma_3/2\pi = \SI{49(2)}{\hertz}$ (quality factor $Q_3=\SI{4.1(2)e3}{}$) of membrane 3.} 
	\label{fig:app:A_lorentzian} 
\end{figure}

\section{Characterization of membranes 2 and 3}\label{app:sec:charac2-3}
Here, we characterize the linear and parametric responses of membranes 2 and 3.

\textbf{Linear responses:} To directly read out the motion of membrane $i = \{2,3\}$, we first move the laser to the center of the corresponding membrane. Then, similarly as in the main text, we measure the linear response of membrane $i$ by applying a voltage $U_i = \bar{U}_i + \tilde{U}_{\omega, {i}} \cos(\omega t)$ to the electrode beneath it and by directly measuring its displacement amplitude $|x_i(\omega)|$ and phase $\mathrm{arg}\{x_i(\omega)\}$ at that same frequency. We detune the membranes which are close in frequency by applying a DC voltage to them. We present this measurement for membrane 2 in Fig.~\ref{fig:app:A_lorentzian}(a), and for membrane 3 in Fig.~\ref{fig:app:A_lorentzian}(b). From the resonance fit, we extract for membrane 2 a frequency $\Omega_2/2\pi\approx\SI{203.40}{\kilo\hertz}$ and a damping rate $\Gamma_2/2\pi = \SI{39.2(6)}{\hertz}$, yielding a quality factor $Q_2=\SI{5.19(8)e3}{}$. For membrane 3, we extract a frequency $\Omega_3/2\pi\approx\SI{203.37}{\kilo\hertz}$ and a damping rate $\Gamma_3/2\pi = \SI{49(2)}{\hertz}$, yielding a quality factor $Q_3=\SI{4.1(2)e3}{}$. The  uncertainty on the fitted values of the frequencies are on the order of $\sim \SI{10}{\milli\hertz}$, which is much smaller than the $\sim \SI{10}{\hertz}$ drifts we typically see in our experiment (for a fixed laser power and position).

\textbf{Parametric responses:} To measure the parametric response of membrane $i = \{2,3\}$, we first position the laser on the corresponding membrane. Then, we apply a voltage $U_i(t) = \bar{U}_i+ \tilde{U}_{2\omega, {i}}\cos(2\omega t)$, and read out the response $x_i(\omega)$ at frequency $\omega$ while sweeping $\omega$ from high to low frequencies. Doing so for a fixed parametric drive strength $\lambda \propto \tilde{U}_{2\omega, {i}}$, we obtain  the top panels in Fig.~\ref{app:fig:A_single_tongues}(i) and Fig.~\ref{app:fig:A_single_tongues}(ii), for membrane 2 and 3, respectively. Note that the readout in this measurement at large parametric drive is highly nonlinear. This is because the amplitude of the motion $x_i$ is enough to bring the membrane motion over a minimum or maximum of the interference fringe, i.e. $|x_i| \gtrsim \SI{1550}{nm}/8$. When looking from high to low frequencies at the parametric sweeps at large parametric drives, e.g. top panels of Fig.~\ref{app:fig:A_single_tongues}(a), after the ``jump'' in the parametric response, the parametric response  first increases and then decreases instead of simply decreasing due to this optical readout nonlinearity.

Repeating this measurement for various parametric drive strengths $\lambda$, we obtain the Arnold tongues shown in  Fig.~\ref{app:fig:A_single_tongues}(b). The outline of the Arnold tongue is given by~\cite{eichler2023classical}
\begin{equation}
\lambda_i=2 \sqrt{\frac{\Gamma_i^2 \omega^2}{\Omega_i^4}+\left(1-\frac{\Omega_i^2}{\omega^2}\right)^2},
\end{equation}
which can be expressed as
\begin{equation}\label{app:eq:lam_lamth}
\frac{\lambda_i}{\lambda_\mathrm{th}}= \sqrt{\frac{\omega^2}{\Omega_i^2}+\left(1-\frac{\omega^2}{\Omega_i^2}\right)^2\frac{\Omega_i^2}{\Gamma_i^2}}
\end{equation}
with a parametric threshold $\lambda_{\mathrm{th},i} =2/Q_i=  2\Gamma_i/\Omega_i$.

\begin{figure}[t]
	\centering
	\includegraphics[width=1\columnwidth]{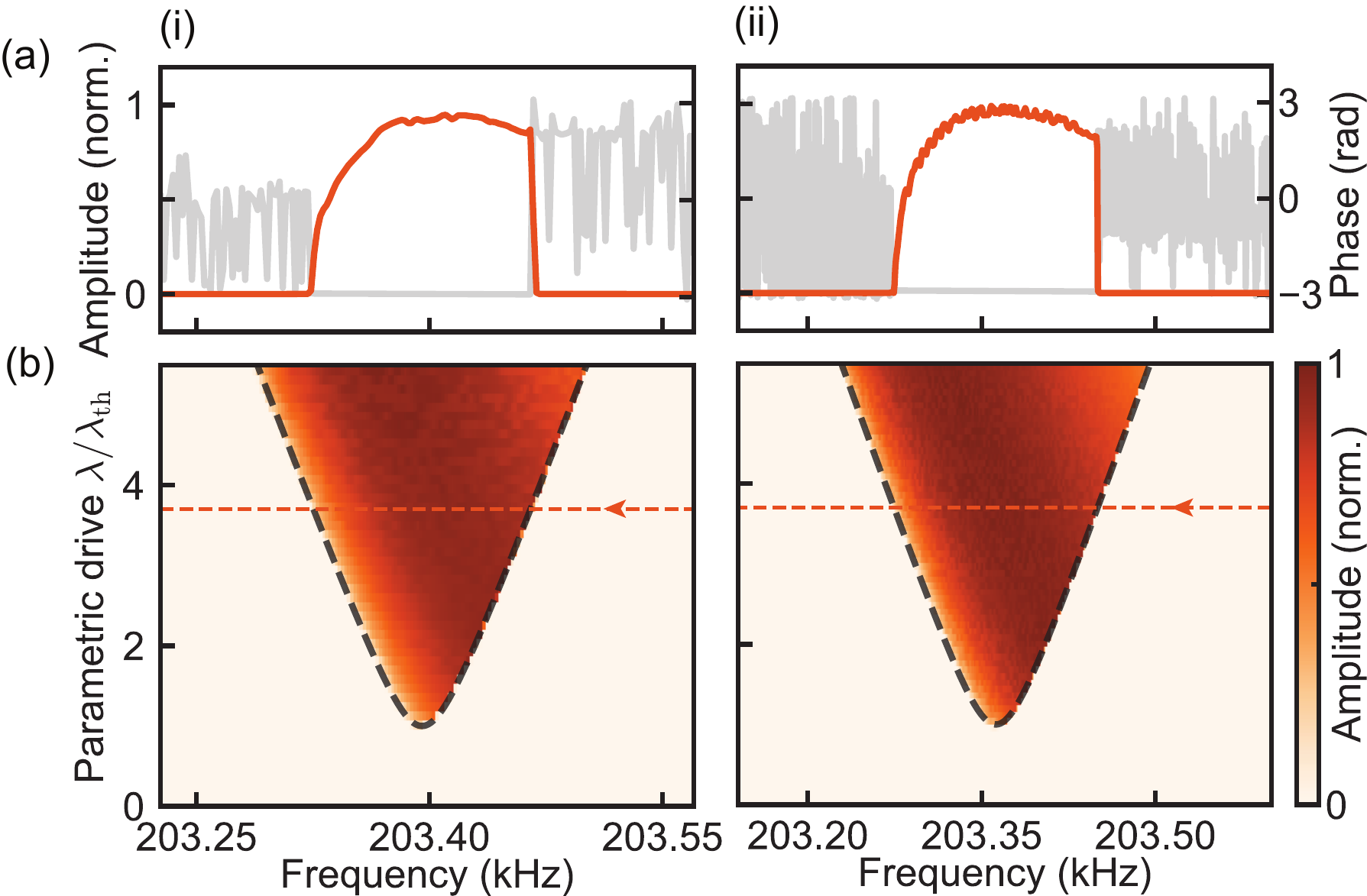}
	\caption{Parametric sweeps of  (i)~membrane 2 and (ii)~membrane 3.  (a)~ Amplitude (orange) and phase (grey) responses for a single parametric sweeps with $\lambda/\lambda_\mathrm{th} = 3.7$. (b)~Arnold tongues obtained by repeating the parametric sweeps for various $\lambda$. The black dashed outlines is the fit of Eq.~\eqref{app:eq:lam_lamth} to our data, with  damping values obtained from the linear responses, see Fig.~\ref{fig:app:A_lorentzian}, and with the frequencies as the only fit parameter, to account for the added DC voltage for the parametric sweep measurements and frequency fluctuations. The sweeps in (a) are marked by dashed orange lines.}
	\label{app:fig:A_single_tongues} 
\end{figure}

\section{Normalized displacement amplitudes}\label{app:sec:double-interferometer}
In the main text, all measured amplitudes $|x_i|$ are presented normalized. The laser light not only reflects from the membrane, but also from the PCB. This forms two low-finesse interferometers, akin to a  membrane inside an optical cavity~\cite{Thompson2008Strong, Jayich2008Dispersive, Dumont2019Flexure}: (1) between the GRIN lens and the membrane, and (2) between the membrane and the PCB. To calibrate for this displacement, we would have to displace the membrane by a quarter of the laser wavelength, which we cannot do in our current setup. In future experiments, we will add a piezoelectric actuator to move the membrane. Alternatively, thermomechanical motion could be used for calibration, but we cannot assume thermal equilibrium due to noise in our voltage source.

\begin{figure}[t]
	\centering
	\includegraphics[width=0.95\columnwidth]{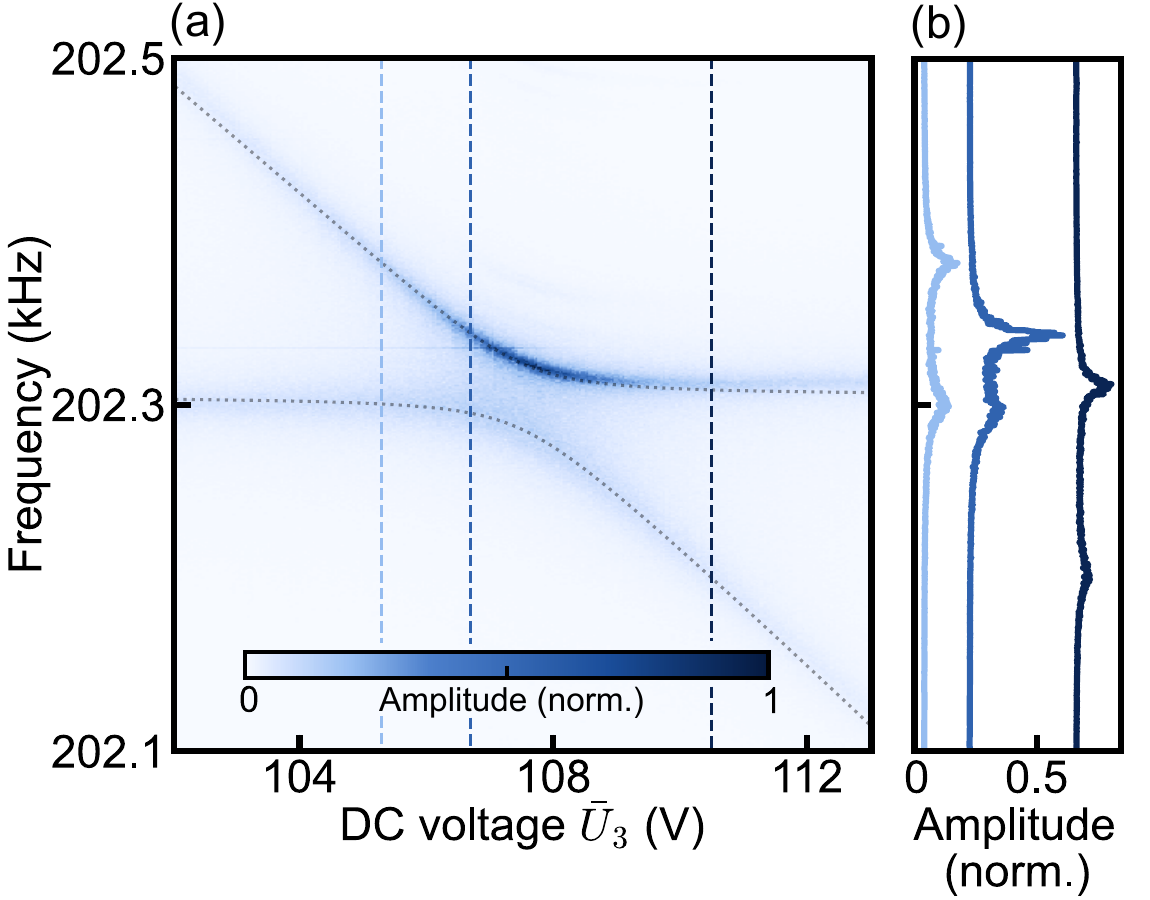}
	\caption{Avoided crossing between  membranes 1 and 3. (a)~Displacement amplitude spectral density of membrane 1 as a function of frequency and $\bar{U}_3$ applied to membrane 3, with fixed $\bar{U}_1 = \SI{34.1}{\volt}$. Dotted lines indicate fits of $\Omega_\pm/2\pi$, using Eq.~\eqref{eq:omega_-+}, yielding coupling $J_{13}/2\pi = \SI{2.77(3)}{\kilo\hertz}$ and an  avoided crossing  gap $\Delta_{13}/2\pi=\SI{37.9 \pm 0.4}{\Hz}$. (b)~Displacement ASD line cuts for different values of $\bar{U}_3$, corresponding to the vertical dashed lines in panel (a).  
    }
	\label{fig:A_avoided_crossing} 
\end{figure}
\section{Avoided crossing model}\label{app:sec:avoided-crossing}
In this appendix, we derive the eigenmodes of two coupled membranes. 
Starting with the coupled equation of motions given by Eq.~\eqref{eq:EOM}, and assuming that the membranes have the same effective mass $m_i=m_j =m$ and symmetric  coupling $J_{ij} = J_{ji}0$, we have the oscillator equations of motion
\begin{align}
\ddot{x}_i+\Omega_i^2 x_i + J_{ij}^2x_j &= 0 , \\
\ddot{x}_j+\Omega_j^2 x_j + J_{ij}^2x_i &= 0 .
\label{App:eq:EOM}
\end{align}
Here, we set $\beta x_{i}^3= \beta x_{j}^3= 0$ and $\Gamma_{i} =\Gamma_{j} =0$ since we are only interested in the eigenmodes frequencies and in their composition in the linear regime. Defining the displacement vector, $\mathbf{x} = (x_i, x_j)^\mathrm{T}$, with $\mathrm{T}$ denoting transpose, this equation can be rewritten in a compact form as
\begin{figure*}[t]
	\centering
	\includegraphics[width=0.8\textwidth]{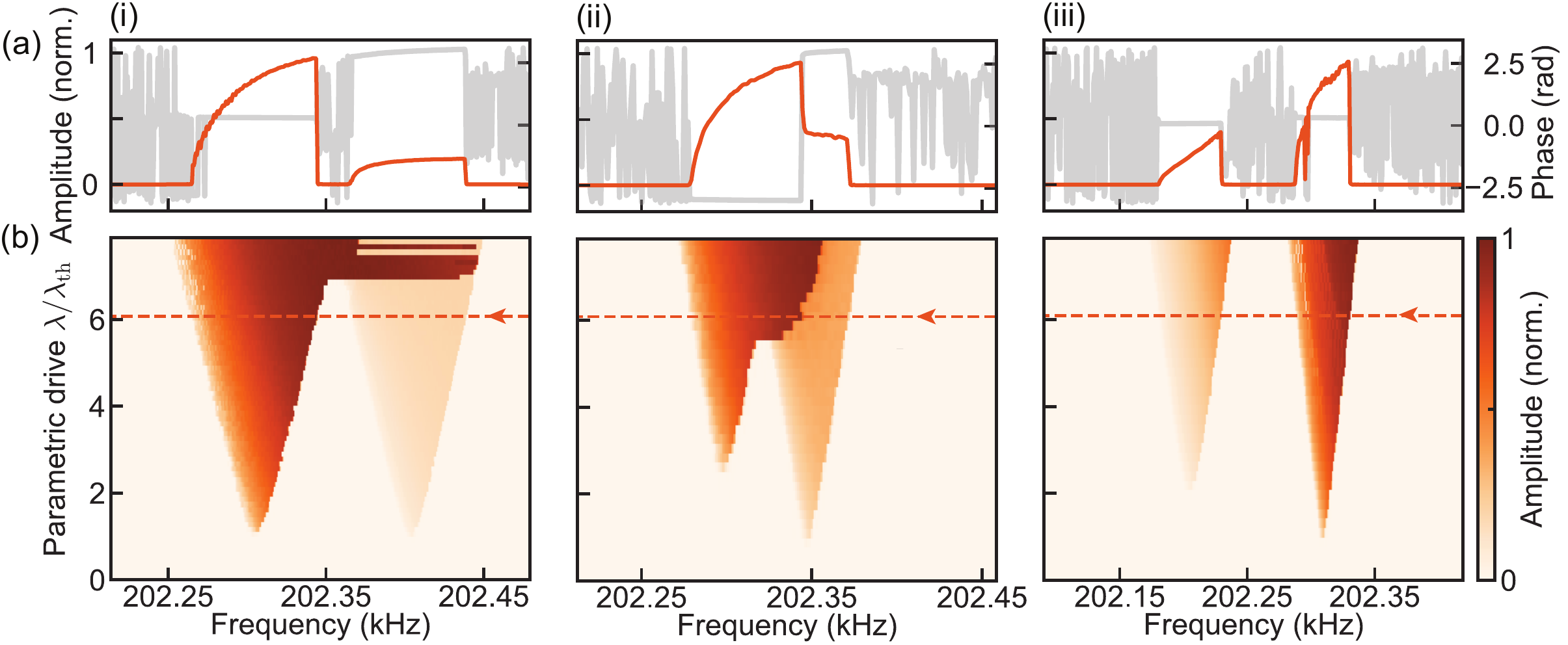}
	\caption{Coupled parametric response of membranes 1 and 3. Columns (i)-(iii) correspond to three different values of $\bar{U}_3$, cf. vertical lines in Fig.~\ref{fig:A_avoided_crossing}. (a)~Parametric amplitude (orange) and phase (grey) responses measured on membrane 1 with $\lambda/\lambda_\mathrm{th} = 6.1$ and $\bar{U}_1=\SI{34.1}{\volt}$. The frequency was swept from high to low values. (b)~Response to parametric driving of the coupled membranes for various $\lambda/\lambda_\mathrm{th}$. The sweeps in (a) were taken along the dashed orange lines.}
	\label{fig:A_coupled_tongues} 
\end{figure*}
\begin{align}
    ( \mathbf{A}-\omega^2\mathbf{I}_2)\mathbf{x} =0,
\end{align}
where ${I}_2 = \mathrm{diag}(1,1)$ is the two-by-two identity matrix, and
\begin{align}
    \mathbf{A} \equiv \begin{pmatrix} \Omega_i^2 & J_{ij}^2\\
J_{ji}^2 & \Omega_j^2 
\end{pmatrix}.
\end{align}
Diagonalizing the matrix $\mathbf{A}$, we obtain the eigenvalues 
\begin{align}\label{app:eq:omega_-+}
    \Omega_\pm^2 = \frac{\Omega_i^2 +\Omega_j^2}{2} \pm \frac{\sqrt{ (\Omega_i^2-\Omega_j^2)^2 + 4J_{ij}^4}}{2},
\end{align}
with $\Omega_\pm$ corresponding to the frequencies of the eigenmodes, which have associated eigenvectors
\begin{align}
    \mathbf{x}_\pm = \pm\frac{1}{\sqrt{( \Omega^2_\pm - \Omega_2^2)^2 + J_{ij}^4}} \begin{pmatrix}
        \Omega^2_\pm - \Omega_2^2\\
        J_{ij}^2
    \end{pmatrix} .
\end{align}

We find the normal mode frequencies at the avoided crossing, i.e. where $\Omega_i = \Omega_j$, 
\begin{align}
    \Omega_\pm = \sqrt{\Omega_i^2 \pm J_{ij}^2} \approx \Omega_1 \pm \frac{J_{ij}^2}{2\Omega_i},
\end{align}
in the limit of $\Omega_i\gg J_{ij}$. This leads to a normal-mode splitting
\begin{align}
    \Delta_{ij} = \Omega_+ - \Omega_- \approx \frac{J_{ij}^2}{\Omega_1}.
\end{align}
We can thus relate the fits of $\Omega_\pm$, which yields $J_{ij}$, to the normal mode splitting.

At the crossing point, the eigenvectors are
\begin{align}
    x_\pm =  x_1 \pm x_2
\end{align}
with $x_+$ ($x_-$) corresponding to the symmetric (antisymmetric) mode with higher (lower) frequency.  Note that a minus sign in front of coupling term $ J_{ij}^2 x_j$ in Eq.~\eqref{eq:EOM} would instead translate to the higher frequency mode being antisymmetric. We verified with a wide-field (illuminating both membranes simultaneously) stroboscopic interferometer that the higher frequency mode is symmetric, confirming the sign we use in Eq.~\eqref{eq:EOM}.

\section{Coupling between membranes 1 and 3}\label{app:sec:coupling1-3}
In the main text, we demonstrate an avoided crossing between nearest-neighbor membranes 1 and 2, and their coupled parametric responses when tuning their frequencies close to each other. Here, we repeat these measurements for membranes 1 and 3, which are diagonally connected.

\textbf{Avoided crossing:} As described in the main text, we measure the avoided crossing between membrane 1 and 3 by reading out membrane 1. We apply a voltage $\bar{U}_1 = \SI{34.1}{\volt}$ to membrane 1 and tune the frequency of membrane 3 by varying $\bar{U}_3$, see  Fig.~\ref{fig:A_avoided_crossing}. Fitting Eq.~\eqref{eq:omega_-+} to our data, we obtain coupling $J_{13}/2\pi = \SI{2.77(3)}{\kilo\hertz}$, leading to normal mode splitting $\Delta_{13}/2\pi=\SI{37.9 \pm 0.4}{\Hz}$. As expected, this splitting is smaller than between membranes 1 and 2 since membranes 1 and 3 are further apart.

\textbf{Tuning of Arnold tongues:} We now measure the coupled parametric responses of membranes 1 and 3, see Fig.~\ref{fig:A_coupled_tongues}. We apply $\bar{U}_1 = \SI{34.1}{\volt}$ to membrane 1, and various $\bar{U}_3$ to membrane 3 corresponding to the vertical lines in Fig,~\ref{fig:A_avoided_crossing}. We apply a parametric drive $\lambda$ to each membrane at frequency $2\omega$ and read out the displacement $x_1$ of membrane 1 at $\omega$, while we sweep $\omega$ from high to low frequencies. The same qualitative behavior as described in the main text can be observed.

\section{Matching parametric drive strengths}\label{app:sec:param_matching}
In order to measure coupled Arnold tongues, we need to carefully tune the parametric drive voltage $\tilde{U}_{2\omega, i}$ applied each membrane. Since each membrane has a different DC voltage $\bar{U}_i$ applied to it, and has a different quality factor $Q_i$, they require different AC voltage amplitudes to achieve the same parametric drive strength $\lambda$. This driving further needs to be tuned due to mode hybridization. In practice, we simply attenuate the AC voltage applied to one of the membrane $\tilde{U}_{2\omega, i}$. Driving at different parametric drive strengths results in various artifacts and different phase diagrams. For instance, in Fig.~\ref{fig:A_coupled_tongues}(i)(b), we can observe such artifacts for sweeps around $\lambda/\lambda_\mathrm{th}$ due to a slight discrepancy in parametric drives for each membrane.

\bibliography{biblio.bib}

\end{document}